\documentclass[a4paper,fleqn,usenatbib]{mnras}
\usepackage{graphicx}	
\usepackage{amsmath}	
\usepackage{amssymb}	
\usepackage{comment,orcidlink}

\def\msun{{\rm M_{\odot}}}

\def\be{\begin{equation}}
\def\ee{\end{equation}}
\def\le{{L_{\rm Edd}}}

\def\del#1{{}}

\newcommand\mj{{\,{\rm M}_{\rm J}}}

\newcommand\MSunPerYear{~${\rm M_{\odot}}$~yr$^{-1}$\,}
\newcommand{\bref}[1]{{ #1}}

\newcommand\konkoly{Konkoly Observatory, HUN-REN Research Centre for Astronomy and Earth Sciences, Konkoly-Thege Mikl\'os \'ut 15--17, 1121 Budapest, Hungary}
\newcommand\elte{Institute of Physics and Astronomy, ELTE E\"otv\"os Lor\'and University, P\'azm\'any P\'eter s\'et\'any 1/A, 1117 Budapest, Hungary}

\newcommand\heidelberg{Max Planck Institute for Astronomy, Königstuhl 17, 69117 Heidelberg, Germany}

\title[]{The youngest of hot jupiters in action: episodic accretion outbursts in Gaia20eae.}

\author[Nayakshin et al.]{Sergei Nayakshin$^{1 \orcidlink{0000-0002-6166-2206}}$\thanks{sergei.nayakshin@le.ac.uk}, Fernando Cruz S\'aenz de Miera$^{2,3,4 \orcidlink{0000-0002-4283-2185}}$ and 
\'Agnes K\'osp\'al$^{3,4,5,6 \orcidlink{0000-0001-7157-6275}}$\\
$^{1}$School of Physics and Astronomy, University of
  Leicester, Leicester, LE1 7RH, UK. \\
$^2$ Institut de Recherche en Astrophysique et Planétologie, Université de
Toulouse, UT3-PS, CNRS, CNES, 9 av. du Colonel Roche, \\
31028 Toulouse Cedex 4, France \\
$^3$\konkoly\\
$^4$ CSFK, MTA Centre of Excellence, Konkoly-Thege Miklós út 15-17, 1121 Budapest, Hungary\\
$^5$ \elte\\
$^6$ \heidelberg
}

\date{Accepted XXX. Received YYY; in original form ZZZ}

\pubyear{2023}

\begin{document}
\label{firstpage}
\pagerange{\pageref{firstpage}--\pageref{lastpage}}
\maketitle

\begin{abstract}
Recent imaging observations with ALMA and other telescopes found widespread signatures of planet presence in protoplanetary discs at tens of au separations from their host stars. Here we point out that the presence of very massive planets at 0.1 au sized orbits can be deduced for protostars accreting gas at very high rates, when their
discs display powerful Thermal Instability bursts. Earlier work showed that a massive planet modifies the nature of this instability, with outbursts triggered at the outer edge of the deep gap opened by the
planet. We present simulations of this effect, finding two types of TI outbursts: downstream and upstream of the planet, which may or may not be causally connected. We apply our model to the outburst in Gaia20eae. We find that the agreement between the data and our disc thermal instability model is improved if there is a planet of 6 Jupiter masses orbiting the star at 0.062 au separation. Gaia20eae thus becomes the second episodically erupting star, after FU Ori, where
the presence of a massive planet is strongly suspected. Future observations of similar systems will constrain the mode
and the frequency of planet formation in such an early epoch.

\end{abstract}


\section{Introduction}

Ionisation of Hydrogen leads to a well known Thermal Instability \citep[TI;][]{Faulkner83-DNe,Meyer84-CVs} in accretion discs, during which the inner disc switches between long periods of ``quiescence" when the disc is neutral (cold), and  shorter outbursts, when the disc is ionised (hot). With outbursts occurring multiple times a year in each of over 200 systems known, TI is the accepted theory of episodic accretion outbursts in dwarf novae \citep[DNe;][]{Dubus18_DIM_DNe,Hameury-20-review} and low mass X-ray binaries \cite[LMXRBs;][]{Coriat12_DIM_LMXRB}. These outbursts are unique in setting tight constraints on the usually very poorly known disc viscosity, indicating that $\alpha \sim 0.01$ ($\sim 0.1$) when H is the neutral (ionised) in the disc \citep{Smak84-dwarf-novae-review,Lasota01-Review}. Recent 3D radiation and \bref{ideal} magneto hydrodynamics (MHD) simulations of discs \citep[e.g.,][]{Hirose15,Coleman16-TI,Scepi-18-TI-alpha} reproduce such viscosity from first principles. 

Cycles of low and high $\dot M$ are observed in accreting protostars as well, with the most powerful of them called  FU Ori type \citep[FUORs;][]{Herbig-77-FUOR-EXOR,HartmannK85-FUORs}. These outbursts are however disappointingly rare, with just a few known pre-2000. All of these outbursts also appeared unhelpfully long, e.g., $\gtrsim 10^2$ years \citep{HK96}, preventing us from knowing how they end. The 
TI theory, re-scaled from stellar binaries to YSO, matches many of the observed ``classic FUOR" characteristics \citep{Clarke-90-FUOR,Bell94,Clarke05-FUORs} but only if $\alpha$ is two orders of magnitude lower than in DNe/LMXRBs. \bref{While lower viscosity is plausible in quiescent cold discs due to non-ideal MHD effects \citep[e.g.,][]{Gammie96,Gressel15-disk-winds}}, this does not appear physically justifiable in outburst, prompting many to question the relevance of TI scenario for FUORs \citep[e.g.,][]{Zhu09-FUOri-obs,Armitage15-review}.

Recent wide sky surveys uncovered a number of outbursts FUOR-like in luminosity but only $\sim 1$ to a few years in duration. \cite{Nayakshin-24-classic-TI} (N24 hereafter) showed that these short outbursts are naturally produced by TI with the same viscosity as in DNe/LMXRBs,  and presented statistical arguments that TI operates in the whole episodic accretors sample. However, N24 noted that additional physics must be present to (a) explain the long classic FUORs, and (b) account for suppression of {\em most} short TI outbursts since  VVV/VVVX survey shows them far less frequent than theory predicts \citep{Contreras-Pena-24}. This additional physics could be other disc instabilities \citep[e.g.,][]{ArmitageEtal01}, stellar mass perturbers \citep[e.g.,][]{Vorobyov-21-flyby-FUORs}, or massive gas planets \citep[][VB06 hereafter]{VB06}. 

Studying TI in conjunction with other physics of YSO discs is natural \citep[e.g.,][]{Pavlyuchenkov-23-TI}. TI operates in the inner $\sim 0.1$~au. \bref{For example, in the \cite{ArmitageEtal01} scenario an outburst is triggered at  $\sim 1$~au, prompting mass transfer towards the star  at high rate \citep[e.g.,][]{Bae14-MRI-2D,Martin14-DZ-MRI}. As the material propagates  through the innermost 0.1~au, TI-outbursts are triggered \citep[e.g., Appendix A1 in][]{Nayakshin23-FUORi-2}.} If future work cements TI status as the first piece of the FUOR puzzle in place, then, following the lead of \cite{Scepi-19-DNe} in DNe, we can use TI to learn what else takes place in FUOR discs.

\bref{In this spirit, and} motivated by the recently discovered {\em short} episodic accretor Gaia20eae \citep{CruzSaenz-22-Gaia20eae}, here we study the model of \cite{LodatoClarke04} (LC04 hereafter) for FUORs. They showed that a gap-opening (massive) planet migrating through the disc modifies TI outbursts significantly. While at separation $a\gtrsim 1$~au, the planet opens a practically impenetrable gap, and the disc downstream of the planet drains onto the star.  The disc upstream of the planet, on the contrary, banks up and swells in mass. When the planet migrates to $a\sim 0.1$~au, TI outbursts begin at the gap outer edge, and propagate outside-in rather than inside-out as in classic TI outbursts. LC04 used the same very low disc viscosity as \cite{Bell94} to obtain 100-year long outbursts\footnote{\cite{Nayakshin-23-FUOR,Nayakshin23-FUORi-2} argue that a gas giant planet that is losing mass via Extreme Evaporation during a TI outburst is a promising model for FU Ori. Here we assume that the planet has a constant mass, which is reasonable for a denser planet, e.g., a planet made by Core Accretion rather than Gravitational Instability.}. Here we use \bref{DNe-like } viscosity values instead, and argue that effects described in LC04 are in action in Gaia20eae.

\section{Methods and a fiducial case}\label{sec:methods}

Our 1D time-dependent planet-in-disc code was described in detail in \cite{Nayakshin-23-FUOR} and in N24. Briefly, our numerical approach to evolving the disc surface density, $\Sigma$, is essentially identical to that of LC04, whereas our energy equation approach is more closely related to \cite{Bell94}. We solve for $\Sigma$ via
\begin{equation}
    \frac{\partial\Sigma}{\partial t} = \frac{3}{R} \frac{\partial}{\partial R} \left[ R^{1/2} \frac{\partial}{\partial R} \left(R^{1/2}\nu \Sigma\right) \right] -
    \frac{1}{R} \frac{\partial}{\partial R} \left(2\Omega^{-1} \lambda \Sigma\right) \;,
\label{dSigma_dt}
\end{equation}
where $\nu=\alpha c_{\rm{s}} H$ is the kinematic viscosity \citep{Shakura73}, $c_{\rm{s}}$ and $H$ are the disc midplane sound speed and vertical scale height, respectively; and $\Omega = (GM_*/R^3)^{1/2}$ is the Keplerian angular frequency. The last term in eq. \ref{dSigma_dt} describes the angular momentum exchange between the disc and the planet via tidal torques; $\lambda$ is specific torque on the disc material. The planet migration rate is set by the angular momentum conservation of the system.

The equations for the evolution of the disc central temperature, $T_{\rm c}$, are given in N24. \bref{Aside from disc viscosity,} another distinction from LC04 is the disc heating due to dissipation of the tidal torque from the planet. This \bref{heating} term, $D_{\rm tid}(R)$, eq. 13 in \cite{LodatoEtal09}, is only significant in  vicinity of the planet. Performing Taylor series expansion of $D_{\rm tid}(R)$ at $R\approx a$ and for $|\Delta R| = |R-a| > \Delta R_{\rm sm}$, we obtain
\begin{equation}
    D_{\rm tid}  \approx \frac{3}{4}q^2\Omega^3 R^2 \Sigma \left(\frac{a}{|\Delta R|}\right)^3 \;,
    \label{Dtid}
\end{equation}
where $a$ is planet-star separation, $q=M_{\rm p}/M_*$, $\Delta R_{\rm sm} = \max(H, R_H)$ is the distance within which both $\lambda$ and $D_{\rm tid}$ are smoothed to go to zero at $\Delta R=0$ (see LC04). One can show that $D_{\rm tid}$ dominates over the viscous dissipation for
\begin{equation}
    \frac{|\Delta R|}{a} \leq  \left(\frac{2}{3}\right)^{1/3} \frac{q^{2/3}}{(\alpha h^2)^{1/3}} \;,
    \label{Dtid_dimensionless}
\end{equation}
where $h=H/R$. For $q=M_{\rm p}/M_*= 5\times 10^{-3}$, eq. \ref{Dtid_dimensionless} yields $\approx 0.25$ in outburst, but can be as large as $0.7$ in quiescence. As in N24 we compute disc spectra by integrating over the disc, assuming the local black-body emission spectrum with a time-dependent  non power-law $T_{\rm eff}(R)$, e.g., see Fig. \ref{fig:Pre-burst}c.


We start with a planet of mass $M_{\rm p} =5\mj$ and initial separation $a=1$~au inserted into a disc with cold and hot viscosity parameters $\alpha_{\rm c} = 0.02$, $\alpha_{\rm h} = 0.1$, and the mass feeding rate of the inner disc $\dot M_{\rm feed} = 7\times 10^{-7}$\MSunPerYear, \bref{added in a narrow ring close to the outer boundary of the grid}. We take $M_*=1.15 \msun$ for Gaia20eae \citep{CruzSaenz-22-Gaia20eae}.

Even though our $\alpha$ values are two orders of magnitude larger than that of LC04, our models reproduce most of their results. Note that since $\dot M  = 3\pi \nu \Sigma$ in an unperturbed planet-free disc at $R\gg R_*$, our disc is $\sim 100$ times lighter at the same $\dot M$. Thus the planet completely dominates system evolution in the inner few au, even more so here than in LC04.

\begin{figure}
\includegraphics[width=0.95\columnwidth]{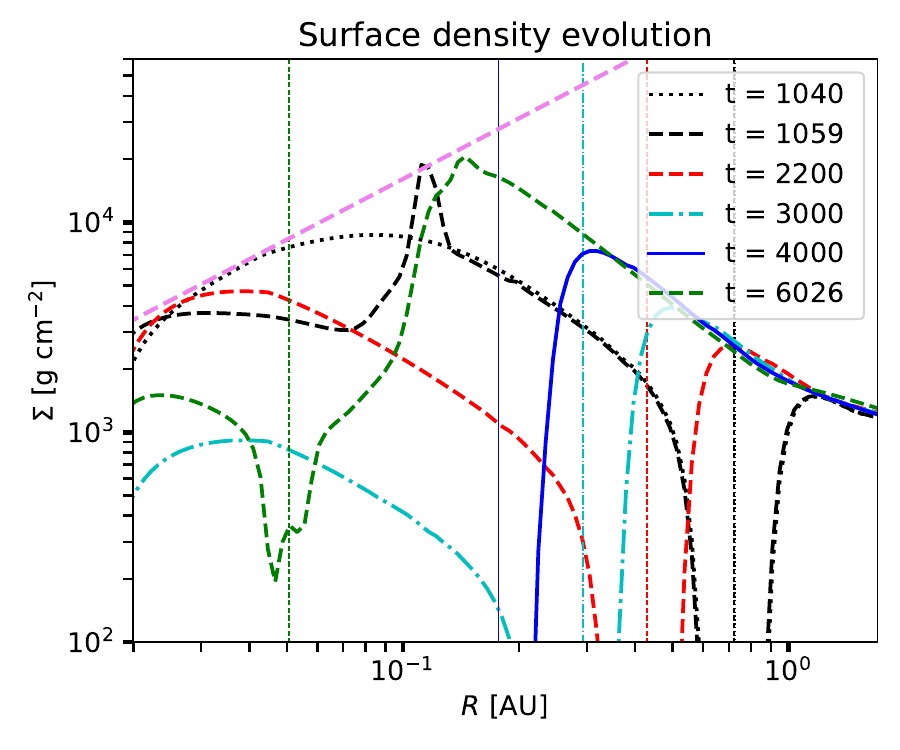}
\caption{Disc $\Sigma(R)$ for a selection of time moments in the fiducial case. Vertical lines mark the respective position of the planet. After the planet is introduced into the disc, it starves the disc downstream of matter, opening a complete inner hole. No TI outbursts occur in that configuration. However, when the planet migrates to $\sim 0.08$~au, TI outburst restart, but this time they are triggered upstream of the planet.}
\label{fig:Sigma_only}
\end{figure}

Fig. \ref{fig:Sigma_only} shows snapshots of $\Sigma(R)$ at specific times  in disc evolution. The violet power-law line shows $\Sigma_{\rm crit}$, termed $\Sigma_{\rm A}$ in Fig. 1 in LC04. Where $\Sigma > \Sigma_{\rm crit}$, a TI outburst is triggered. As in LC04, when the planet is introduced into the disc, it cuts a deep gap. Initially the inner disc is largely oblivious to this and continues nearly classic planet-free TI outbursts, e.g., they are triggered at $\sim 2$ stellar radii and propagate outward (see Fig. 4 in N24). In Fig. \ref{fig:Sigma_only}, snapshots at $t=1040$ and $1059$ show $\Sigma$ just before and at the very end of an outburst.

However, with time the inner disc drains onto the star. By $t\approx 2200$~years, when $a\approx 0.4$~au, the downstream $\Sigma$ never reaches $\Sigma_{\rm crit}$, and TI outbursts stop. The planet is then slowly pushed towards the star while the downstream disc is completely emptied so an inner hole is formed (e.g., see $t=4,000$ in Fig. \ref{fig:Sigma_only}a).

The disc upstream of the planet, on the contrary, banks up and swells in mass. As in LC04, when the planet migrates to the inner $\sim 0.1$~au, the peak in $\Sigma$ just upstream of the planet edges very close to $\Sigma_{\rm crit}$, and the outbursts can restart there. The  outbursts are triggered upstream of the planet and propagate in. The last snapshot (dashed green) in Fig. \ref{fig:Sigma_only} shows $\Sigma$ at the end of such an outburst.

\begin{figure}
\includegraphics[width=0.95\columnwidth]{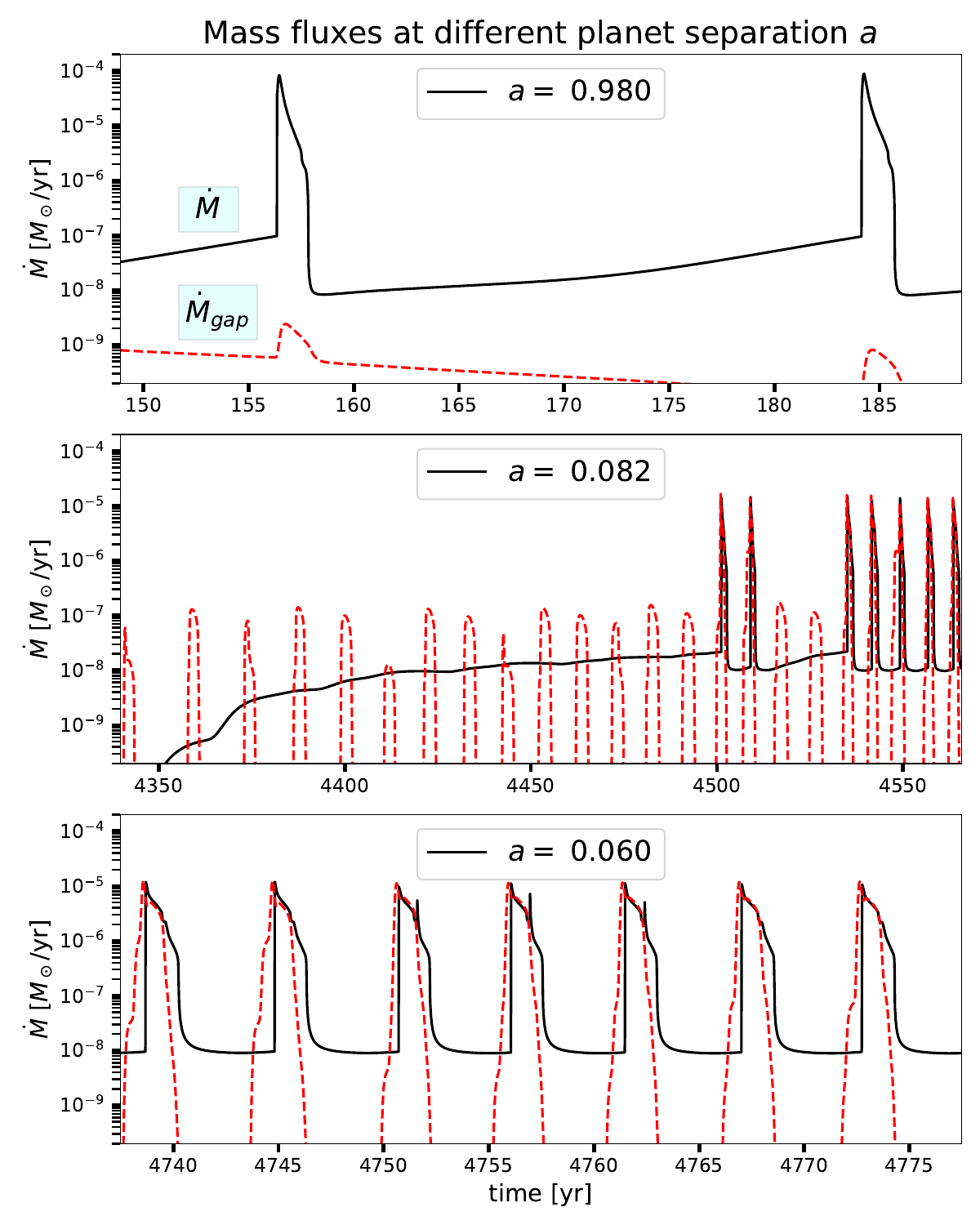}
\caption{Accretion rate onto the star (black) and through the gap region (red) vs time at three moments in the fiducial simulation. The large peaks in $\dot M$ and $\dot M_{\rm gap}$ are TI bursts downstream and upstream of the planet, respectively. The planet gap region is impenetrable barrier in the top panel but becomes leaky when the planet migrates sufficiently in (the two bottom panels). Note that an upstream burst initiates each of the downstream bursts in the bottom panel. The time span in the middle panel  is far longer than in the two other panels. See \S \ref{sec:methods} for detail.}
\label{fig:a-Mdot-time}
\end{figure}


Since we include $D_{\rm tid}$ here, and due to a much higher $\alpha$ than in LC04, we find additional observable signatures of upstream TI outbursts. Fig. \ref{fig:a-Mdot-time} shows mass fluxes in two important locations in the disc: $\dot M$ in the accretion rate on the star,  while $\dot M_{\rm gap} = -2\pi \Sigma(R_a) v_R(R_a)$ is the mass flux through $R_a$, the radial bin containing the planet at that moment. Here $v_R$ is the radial velocity of the gas in that bin. These mass fluxes are shown at three moments illuminating the planet-disc system evolution best. The central legend in each panel shows the respective planet location.

In the top panel in Fig. \ref{fig:a-Mdot-time}, soon after the planet has been inserted into the disc, $\dot M$ is characteristic of the classic TI outbursts, whereas $\dot M_{\rm gap}$ plunges to negligible levels as the gap is being excavated. As explained above, this cuts off the innermost disc of further fuel supply, and the downstream outbursts stop when the innermost disc runs out of matter; $\dot M$ then plunges to very small levels too.

This goes on until the planet is at $a\approx 0.08a$~au, when the disc just upstream of the planet is hot enough to experience local ``upstream bursts", discussed more fully in \S \ref{sec:Gaia20eae}. In these bursts the disc at the outer edge of the gap can heat up to $T\gtrsim 10^4$~K (Fig. \ref{fig:Pre-burst}b), and the gap region becomes leaky. The material then rushes through the gap into the downstream disc (cf. the middle panel in Fig. \ref{fig:a-Mdot-time}), but initially no TI activity takes place there because $\Sigma < \Sigma_{\rm crit}$. After a succession of ``gulps" through which the downstream disc receives its mass from upstream of the planet, the downstream bursts return at $t\approx 4500$~years. These bursts are different from the classic planet-free ones, though, because the mass supply into the inner disc is still controlled by the upstream bursts. We believe that LC04 did not see the gulps that fail to propagate a TI burst all the way to the star because their bursts  were dozens of times longer, each bringing much more matter downstream from the planet than ours.

As the planet migrates further in, the upstream bursts become more and more powerful, and are eventually able to push through enough mass in each gulp for the downstream disc to experience a TI burst every time. The upstream and downstream outbursts are then synchronised in the sense that the former always causes the latter (bottom panel in Fig. \ref{fig:a-Mdot-time}).

While investigating parameter space of this scenario we came across outbursts with lightcurves resembling that of Gaia20eae. These outbursts start upstream of the planet, and the planet tends to be at $a < 0.1$~au. 

%


\section{The model for Gaia20eae}\label{sec:Gaia20eae}

To save computational time, our simulations aimed at addressing Gaia20eae start with the planet injected into the disc at $a=0.2$~au at $t=0$. To approximate the complete inner hole that the planet opens in more realistic simulations where it migrates from much further away, we set disc $\Sigma$ to zero within 0.25~au. The planet is then pushed towards the star by the planet-disc interactions. By the time the planet is at $a=0.1$~au the system adjusts to a structure resembling the one we found in longer time simulations discussed in \S \ref{sec:methods}. 


We experimented with several free parameters of our model, i.e., $M_{\rm p}$, viscosity parameters $\alpha_{\rm c}$ and $\alpha_{\rm h}$, and $\dot M_{\rm feed}$ to find a combination of parameters that reproduce Gaia20eae outburst as closely as possible. Here we focus only on the ``best case scenario" model, for which $M_{\rm p} =6\mj$, $\alpha_{\rm c} = 0.04$, $\alpha_{\rm h} = 0.1$, and $\dot M_{\rm feed} = 7\times 10^{-7}$\MSunPerYear.

\begin{figure*}
\includegraphics[width=0.98\columnwidth]{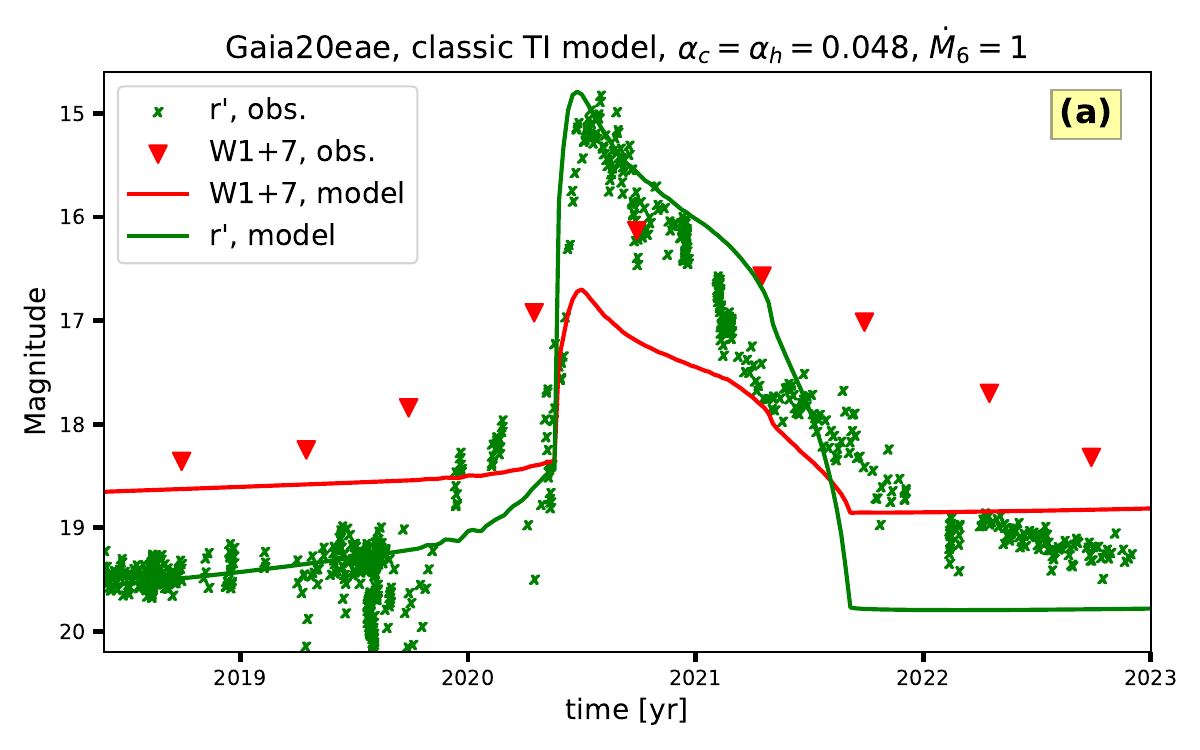}
\includegraphics[width=0.98\columnwidth]{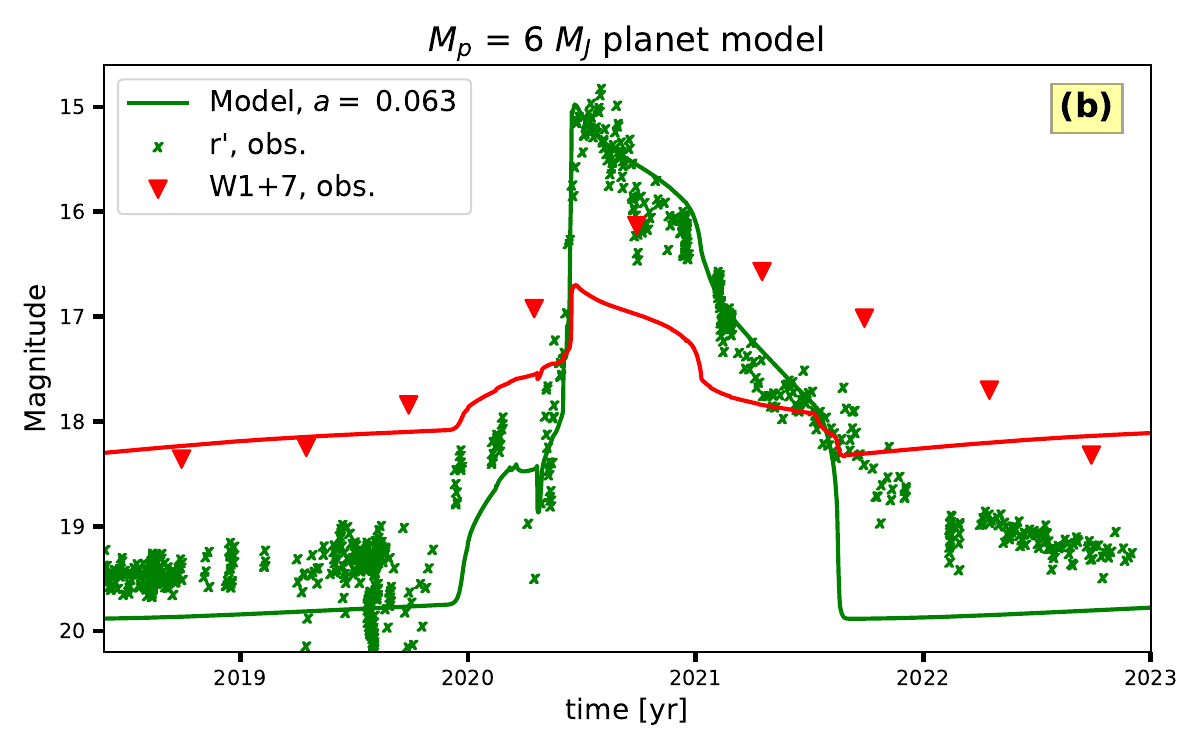}
\caption{Observed photometry in two wavebands (symbols; WISE W1 and Sloan r' bands) and model (lines) for the outburst in Gaia20eae. Panel (a): the classic TI model from Fig. 8 in Nayakshin et al (2024); Panel (b): same but with an embedded planet, this paper.  Two moments in time are shown, with the respective planet locations listed in the legend. The banking up of material upstream of the planet explains the gradual rise in W1, and the pre-burst in r' at early 2020. The drops in the lightcurve in the r' band seen at 2020.3 and 2021.0 correspond to the times when the flow bridges the gap, and when it re-opens, respectively.}
\label{fig:Data_vs_models}
\end{figure*}

Fig. \ref{fig:Data_vs_models} shows the observed lightcurve of Gaia20eae in two bands, W1 and $r'$, with symbols, and two model lightcurves in the same bands. Fig. \ref{fig:Data_vs_models}a shows the model from N24 (their Fig. 8), reproduced here for comparison. This model has no embedded planet. Fig. \ref{fig:Data_vs_models}b is our best case scenario from this paper. The outburst analysed here occurrs when the planet reaches separation $a=0.062$~au.

The presence of the planet improves the agreement between the simulations and the observations in several ways: 
\begin{enumerate}
    \item The disc is brighter in mid-IR before the outburst. This is due to the banking up of material upstream of the planet (e.g., Fig. 3 in LC04). Since local energy liberation is proportional to $\Sigma$, the disc is brighter upstream of the planet than without it. 
    \item There is a small, $\sim 1$ mag, ``precursor burst" seen in $r'$ that begins at $t\approx 2020.0$, several months before the main outburst. The burst ends with a short term drop in $r'$ at $t\approx 2020.3$, just before the rise towards the main burst. These effects are related to the gap region being bridged and are explained together with point (iv) below.
    \item The initial rise to the maximum light in $r'$ is less steep than in the classic TI case. \cite{Nayakshin23-FUORi-2} obtained similar conclusions for the lightcurve rises in the visual B and V bands for the model scenarios (their \S 5.1 and 5.2). Note that this contrasts with LC04 conclusions, which may reflect the large difference in viscosity parameters between the two studies.
    \item There is a $\sim 1$ mag step-like drop in $r'$ at $t\approx 2021$. This drop is caused by the gap in the disc  re-emerging during the cooling phase of the burst. This curtails the spatial extent of the region bright in the $r'$ band.
\end{enumerate}

\begin{figure*}
\includegraphics[width=0.98\textwidth]{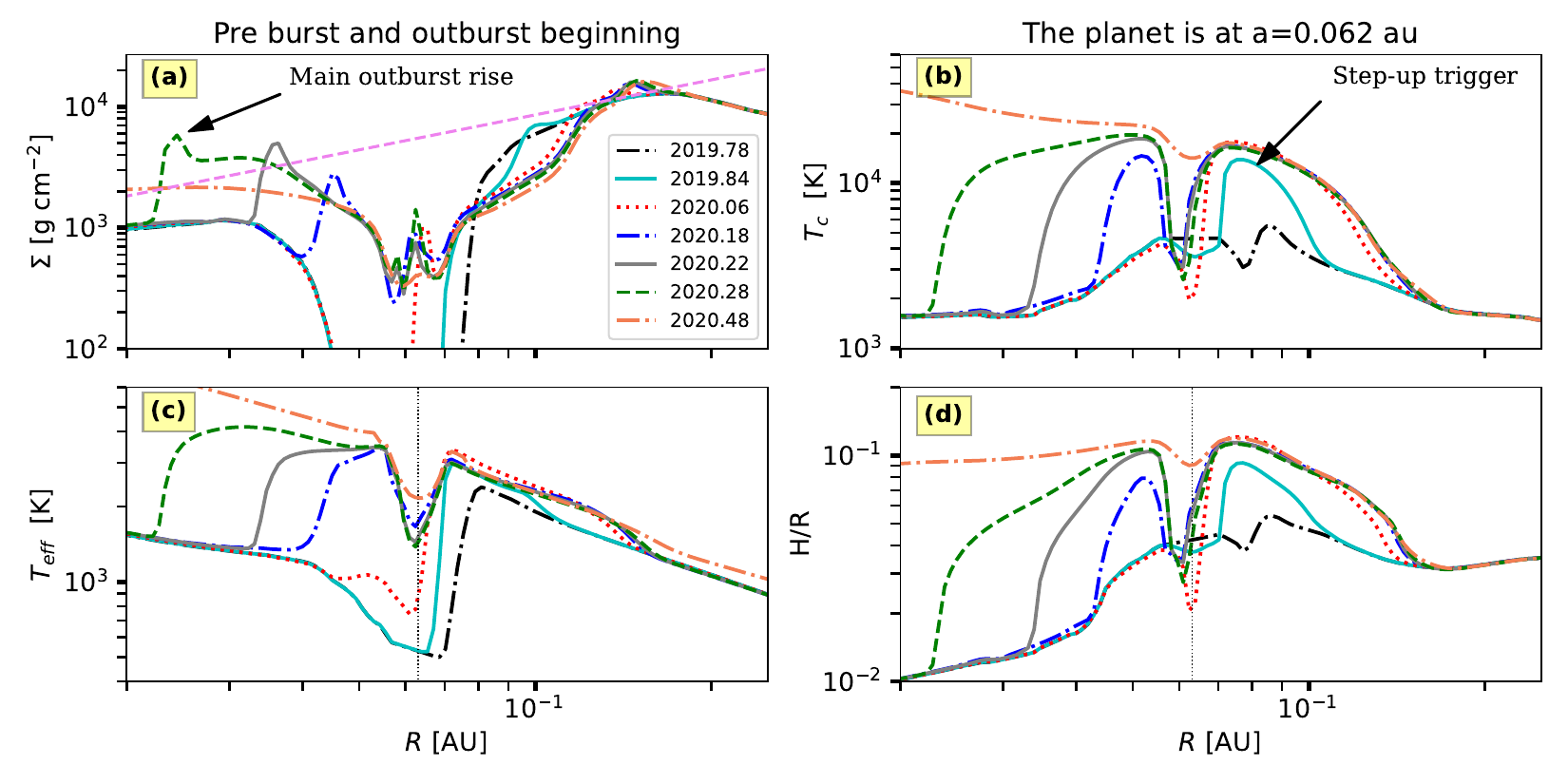}
\caption{Disc evolution trhough the early phases of the outburst shown in the right panel of Fig. \ref{fig:Data_vs_models}. Before the outburst, a deep gap cuts the inner disc off from a steady mass supply from outside. The outburst is triggered just upstream of the planet (cyan curve). As the gas heats up, the material is able to cross the gap and fill the inner disc sufficiently for the outburst to reach all the way to the star.}
\label{fig:Pre-burst}
\end{figure*}

For a deeper discussion of points (ii) and (iv), Fig. \ref{fig:Pre-burst} shows several snapshots of the main disc variables around the time of the precursor burst. As in Fig. \ref{fig:Sigma_only}, when $\Sigma> \Sigma_{\rm crit}$, a classic TI outburst should commence. The precursor burst in Fig. \ref{fig:Data_vs_models}b is however triggered at a much smaller $\Sigma$, at $R\approx 0.075$~au, cf. the cyan curve in Fig. \ref{fig:Pre-burst}b. This happens due to the strong extra heating that the disc receives from the planet via tidal dissipation $D_{\rm tid}$ (eq. \ref{Dtid}), which has a local maximum at the trigger point (eq. \ref{Dtid_dimensionless}). At larger $R$, $D_{\rm tid} \propto \Delta R^{-3}$, so falls very rapidly. $D_{\rm tid}$ also falls strongly within the gap where $\Sigma$ is small. We found that the precursor bursts disappear in our model if we set $D_{\rm tid}=0$.

The precursor bursts are very distinct from the classic TI ones. The latter are triggered very close to the star, and hence lead to an increase in $T_{\rm eff}$ from $\sim 3,000$ K to as much as $\sim 10,000$~K \citep[e.g., Fig. 4 in][]{Nayakshin23-FUORi-2}. This fuels $\sim 5$ mag brightness increase of the disc in the optical bands. The precursor bursts  are triggered on the outer edge of the gap, factor of $\sim 5$ further out in the disc. Therefore, in precursor bursts the local effective temperature cannot increase much. The disc luminosity in the visual does not increase significantly while the ionisation front propagates inward due to the strong advective cooling (cf. Figs. \ref{fig:Pre-burst}b and c).

\begin{figure}
\includegraphics[width=0.98\columnwidth]{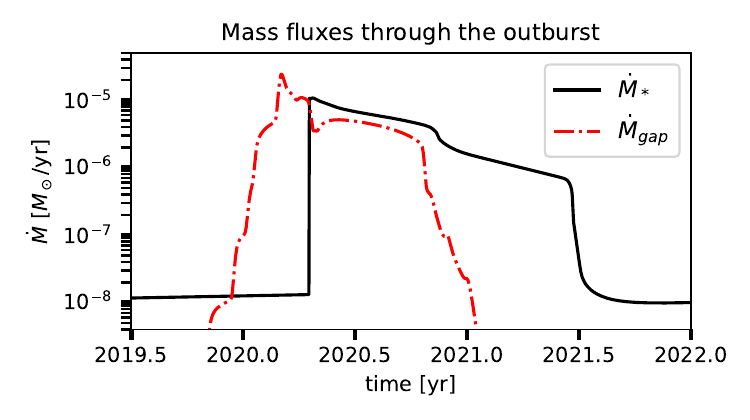}
\caption{Mass accretion rate onto the star (black) and through the planet location (red). The step-up 1 mag burst seen in Fig. \ref{fig:Data_vs_models} at $t\approx 2020$ begins when the gap is bridged by the material crossing it from the exterior to the interior disc. As the disc cools, the gap closes, which induces a step down in $\dot M$ at around $t\approx 2021$.}
\label{fig:Mdot_only}
\end{figure}

During the main outburst the gap region remains somewhat suppressed in $\Sigma$ and cooler than the disc just outside; nevertheless, the material is able to pass through it relatively easily. Fig. \ref{fig:Mdot_only} depicts the mass flow rate $\dot M$ onto the star and through the gap region, $\dot M_{\rm gap}$, vs time. We observe that beginning of the precursor burst coincides material upstrem of the planet starting to flow through the gap region into the inner disc. The main outburst commences when the ionisation front reaches the star and $\dot M$ suddenly skyrockets. Fig. \ref{fig:a-Mdot-time} also shows that the mass flow rate through the gap is comparable to $\dot M$ over much of the 2020. This confirms that the pre-burst phenomenon is crucial to powering the whole of the outburst; it only appears as a small (1 mag) step-up in the $r'$ because the region is not as bright in the optical as the hotter regions downstream of the planet during the main burst.

When the disc starts to cool, the upstream region falls back into quiescence first. This triggers re-emergence of the gap and termination of the gas flow through the gap at $t\approx 2020.8$. The inner disc starves of fuel, which necessitates a drop in $\dot M$ onto the star seen in the black curve in Fig. \ref{fig:Mdot_only},
and a $\sim 1 $ mag drop in r' band shortly thereafter. The inner disc outburst continues until $\approx 2021.5$ when it finally falls into quiescence.

\section{Discussion and Conclusions}\label{sec:Discussion}

Here we confirmed results of \cite{LodatoClarke04} for TI outbursts in discs with massive gap-opening planets, and found new specific signatures of this scenario. Such planets isolate the inner disc from the mass reservoir -- the outer disc -- so the classic inside-out TI outbursts disappear together with the inner disc. When the planet migrates inside $R \sim 0.1$~au, the outbursts restart, but are triggered upstream of the planet and propagate outside-in.

We used a much larger disc viscosity than did LC04, as is consistent with TI in stellar binary systems. Due to this, TI outbursts starting on the gap outer edge are much shorter and transfer much less mass into the inner disc. The mass of the latter hence increases in small ``gulps" (cf. Fig. \ref{fig:a-Mdot-time}). A number of these is needed to refill the inner disc so that TI can restart there. Eventually the upstream and downstream discs come into equilibrium in which every gulp tips the inner disc over the instability threshold, and TI propagates all the way to the star. The outbursts in this case start with a small (in terms of the source magnitude change, say $\Delta r'$) step-up burst, which corresponds to the ``gulp". This is then followed by the downstream burst with a much larger $\Delta r'$.

We applied this scenario to the outburst of Gaia20eae \citep{CruzSaenz-22-Gaia20eae}, finding a somewhat better fit to its lightcurve than without a planet. We note that the fit is far from perfect, but it is probably not realistic to expect anything better with a 1D modelling that employs a simple $\alpha$-prescription for viscosity. Further, some of the disagreements may be expected given missing physics in the model. For example, Fig. 9 in \cite{Bell-99-FUOR-shadow} shows how reprocessing of the radiation in the $R\gtrsim 1$~AU disc makes the disc significantly brighter during FUOR outbursts in IR; we do not include these effects. It is possible that disc irradiation could make our model brighter in W1 band, and make it more consistent with the observations of Gaia20eae.

Additionally, the stellar magnetosphere is neglected in the model \citep[since magnetospheres appear to be crushed by the disc during FUOR outbursts, e.g.,][]{HK96}. However, closer to the end of the outburst the magnetosphere may recover. Preliminary numerical experiments show that this may explain the smooth decay into quiescence in Gaia20eae compared with theoretical model abrupt cutoff in Fig. \ref{fig:Data_vs_models}.
 
Notwithstanding these model shortcomings, our calculations show that the existence of the gap opened by the planet in the disc has specific lightcurve signatures that may help to uncover massive planet presence in FUOR discs. Finally, another possible signature of planet presence is a short time scale quasi-periodic variability in the disc line spectra and photometry, as is observed in FU Ori \citep[e.g.,][]{Herbig03-FUOR,ClarkeArmitage03,PowellEtal12,Siwak21-FUOri-QPOs}. For a planet without mass loss, as considered here, the pumping mechanism of such variability would be planet-disc tidal interactions, especially if the planet gains eccentricity. Future 2D simulations are needed to evaluate such effects better.

\section{Aknowledgement}

We thank Cathie Clarke and Giuseppe Lodato for illuminating discussions of the problem. SN acknowledges the funding from the UK Science and Technologies Facilities Council, grant No. ST/S000453/1. This work was also supported by the NKFIH excellence grant TKP2021-NKTA-64. FCSM received financial support from the European Research Council (ERC)
under the European Union’s Horizon 2020 research and innovation
programme (ERC Starting Grant “Chemtrip”, grant agreement No 949278).

\section{Data availability}

The data obtained in our simulations can be made available on reasonable request to the corresponding author.


\bibliographystyle{mnras}
\bibliography{nayakshin}

\begin{thebibliography}{}
\makeatletter
\relax
\def\mn@urlcharsother{\let\do\@makeother \do\$\do\&\do\#\do\^\do\_\do\%\do\~}
\def\mn@doi{\begingroup\mn@urlcharsother \@ifnextchar [ {\mn@doi@}
  {\mn@doi@[]}}
\def\mn@doi@[#1]#2{\def\@tempa{#1}\ifx\@tempa\@empty \href
  {http://dx.doi.org/#2} {doi:#2}\else \href {http://dx.doi.org/#2} {#1}\fi
  \endgroup}
\def\mn@eprint#1#2{\mn@eprint@#1:#2::\@nil}
\def\mn@eprint@arXiv#1{\href {http://arxiv.org/abs/#1} {{\tt arXiv:#1}}}
\def\mn@eprint@dblp#1{\href {http://dblp.uni-trier.de/rec/bibtex/#1.xml}
  {dblp:#1}}
\def\mn@eprint@#1:#2:#3:#4\@nil{\def\@tempa {#1}\def\@tempb {#2}\def\@tempc
  {#3}\ifx \@tempc \@empty \let \@tempc \@tempb \let \@tempb \@tempa \fi \ifx
  \@tempb \@empty \def\@tempb {arXiv}\fi \@ifundefined
  {mn@eprint@\@tempb}{\@tempb:\@tempc}{\expandafter \expandafter \csname
  mn@eprint@\@tempb\endcsname \expandafter{\@tempc}}}

\bibitem[\protect\citeauthoryear{{Armitage}}{{Armitage}}{2015}]{Armitage15-review}
{Armitage} P.~J.,  2015, arXiv e-prints, \href
  {https://ui.adsabs.harvard.edu/abs/2015arXiv150906382A} {p. arXiv:1509.06382}

\bibitem[\protect\citeauthoryear{{Armitage}, {Livio}  \& {Pringle}}{{Armitage}
  et~al.}{2001}]{ArmitageEtal01}
{Armitage} P.~J.,  {Livio} M.,   {Pringle} J.~E.,  2001, \mn@doi [\mnras]
  {10.1046/j.1365-8711.2001.04356.x}, \href
  {http://adsabs.harvard.edu/abs/2001MNRAS.324..705A} {324, 705}

\bibitem[\protect\citeauthoryear{{Bae}, {Hartmann}, {Zhu}  \& {Nelson}}{{Bae}
  et~al.}{2014}]{Bae14-MRI-2D}
{Bae} J.,  {Hartmann} L.,  {Zhu} Z.,   {Nelson} R.~P.,  2014, \mn@doi [\apj]
  {10.1088/0004-637X/795/1/61}, \href
  {https://ui.adsabs.harvard.edu/abs/2014ApJ...795...61B} {795, 61}

\bibitem[\protect\citeauthoryear{{Bell}}{{Bell}}{1999}]{Bell-99-FUOR-shadow}
{Bell} K.~R.,  1999, \mn@doi [\apj] {10.1086/307988}, \href
  {https://ui.adsabs.harvard.edu/abs/1999ApJ...526..411B} {526, 411}

\bibitem[\protect\citeauthoryear{{Bell} \& {Lin}}{{Bell} \&
  {Lin}}{1994}]{Bell94}
{Bell} K.~R.,  {Lin} D.~N.~C.,  1994, \mn@doi [\apj] {10.1086/174206}, \href
  {http://ukads.nottingham.ac.uk/cgi-bin/nph-bib_query?bibcode=1994ApJ...427..987B&db_key=AST}
  {427, 987}

\bibitem[\protect\citeauthoryear{{Clarke} \& {Armitage}}{{Clarke} \&
  {Armitage}}{2003}]{ClarkeArmitage03}
{Clarke} C.~J.,  {Armitage} P.~J.,  2003, \mn@doi [\mnras]
  {10.1046/j.1365-8711.2003.06983.x}, \href
  {https://ui.adsabs.harvard.edu/abs/2003MNRAS.345..691C} {345, 691}

\bibitem[\protect\citeauthoryear{{Clarke}, {Lin}  \& {Pringle}}{{Clarke}
  et~al.}{1990}]{Clarke-90-FUOR}
{Clarke} C.~J.,  {Lin} D.~N.~C.,   {Pringle} J.~E.,  1990, \mn@doi [\mnras]
  {10.1093/mnras/242.3.439}, \href
  {https://ui.adsabs.harvard.edu/abs/1990MNRAS.242..439C} {242, 439}

\bibitem[\protect\citeauthoryear{{Clarke}, {Lodato}, {Melnikov}  \&
  {Ibrahimov}}{{Clarke} et~al.}{2005}]{Clarke05-FUORs}
{Clarke} C.,  {Lodato} G.,  {Melnikov} S.~Y.,   {Ibrahimov} M.~A.,  2005,
  \mn@doi [\mnras] {10.1111/j.1365-2966.2005.09231.x}, \href
  {https://ui.adsabs.harvard.edu/abs/2005MNRAS.361..942C} {361, 942}

\bibitem[\protect\citeauthoryear{{Coleman}, {Kotko}, {Blaes}, {Lasota}  \&
  {Hirose}}{{Coleman} et~al.}{2016}]{Coleman16-TI}
{Coleman} M.~S.~B.,  {Kotko} I.,  {Blaes} O.,  {Lasota} J.~P.,   {Hirose} S.,
  2016, \mn@doi [\mnras] {10.1093/mnras/stw1908}, \href
  {https://ui.adsabs.harvard.edu/abs/2016MNRAS.462.3710C} {462, 3710}

\bibitem[\protect\citeauthoryear{{Contreras Pe{\~n}a}, {Lucas}, {Guo}  \&
  {Smith}}{{Contreras Pe{\~n}a} et~al.}{2024}]{Contreras-Pena-24}
{Contreras Pe{\~n}a} C.,  {Lucas} P.~W.,  {Guo} Z.,   {Smith} L.,  2024,
  \mn@doi [\mnras] {10.1093/mnras/stad3780}, \href
  {https://ui.adsabs.harvard.edu/abs/2024MNRAS.528.1823C} {528, 1823}

\bibitem[\protect\citeauthoryear{{Coriat}, {Fender}  \& {Dubus}}{{Coriat}
  et~al.}{2012}]{Coriat12_DIM_LMXRB}
{Coriat} M.,  {Fender} R.~P.,   {Dubus} G.,  2012, \mn@doi [\mnras]
  {10.1111/j.1365-2966.2012.21339.x}, \href
  {https://ui.adsabs.harvard.edu/abs/2012MNRAS.424.1991C} {424, 1991}

\bibitem[\protect\citeauthoryear{{Cruz-S{\'a}enz de Miera}
  et~al.,}{{Cruz-S{\'a}enz de Miera} et~al.}{2022}]{CruzSaenz-22-Gaia20eae}
{Cruz-S{\'a}enz de Miera} F.,  et~al., 2022, \mn@doi [\apj]
  {10.3847/1538-4357/ac477f}, \href
  {https://ui.adsabs.harvard.edu/abs/2022ApJ...927..125C} {927, 125}

\bibitem[\protect\citeauthoryear{{Dubus}, {Otulakowska-Hypka}  \&
  {Lasota}}{{Dubus} et~al.}{2018}]{Dubus18_DIM_DNe}
{Dubus} G.,  {Otulakowska-Hypka} M.,   {Lasota} J.-P.,  2018, \mn@doi [\aap]
  {10.1051/0004-6361/201833372}, \href
  {https://ui.adsabs.harvard.edu/abs/2018A&A...617A..26D} {617, A26}

\bibitem[\protect\citeauthoryear{{Faulkner}, {Lin}  \& {Papaloizou}}{{Faulkner}
  et~al.}{1983}]{Faulkner83-DNe}
{Faulkner} J.,  {Lin} D.~N.~C.,   {Papaloizou} J.,  1983, \mn@doi [\mnras]
  {10.1093/mnras/205.2.359}, \href
  {https://ui.adsabs.harvard.edu/abs/1983MNRAS.205..359F} {205, 359}

\bibitem[\protect\citeauthoryear{{Gammie}}{{Gammie}}{1996}]{Gammie96}
{Gammie} C.~F.,  1996, \mn@doi [\apj] {10.1086/176735}, \href
  {http://adsabs.harvard.edu/abs/1996ApJ...457..355G} {457, 355}

\bibitem[\protect\citeauthoryear{{Gressel}, {Turner}, {Nelson}  \&
  {McNally}}{{Gressel} et~al.}{2015}]{Gressel15-disk-winds}
{Gressel} O.,  {Turner} N.~J.,  {Nelson} R.~P.,   {McNally} C.~P.,  2015,
  \mn@doi [\apj] {10.1088/0004-637X/801/2/84}, \href
  {https://ui.adsabs.harvard.edu/abs/2015ApJ...801...84G} {801, 84}

\bibitem[\protect\citeauthoryear{{Hameury}}{{Hameury}}{2020}]{Hameury-20-review}
{Hameury} J.~M.,  2020, \mn@doi [Advances in Space Research]
  {10.1016/j.asr.2019.10.022}, \href
  {https://ui.adsabs.harvard.edu/abs/2020AdSpR..66.1004H} {66, 1004}

\bibitem[\protect\citeauthoryear{{Hartmann} \& {Kenyon}}{{Hartmann} \&
  {Kenyon}}{1985}]{HartmannK85-FUORs}
{Hartmann} L.,  {Kenyon} S.~J.,  1985, \mn@doi [\apj] {10.1086/163713}, \href
  {https://ui.adsabs.harvard.edu/abs/1985ApJ...299..462H} {299, 462}

\bibitem[\protect\citeauthoryear{{Hartmann} \& {Kenyon}}{{Hartmann} \&
  {Kenyon}}{1996}]{HK96}
{Hartmann} L.,  {Kenyon} S.~J.,  1996, \mn@doi [\araa]
  {10.1146/annurev.astro.34.1.207}, \href
  {http://adsabs.harvard.edu/abs/1996ARA%26A..34..207H} {34, 207}

\bibitem[\protect\citeauthoryear{{Herbig}}{{Herbig}}{1977}]{Herbig-77-FUOR-EXOR}
{Herbig} G.~H.,  1977, \mn@doi [\apj] {10.1086/155615}, \href
  {https://ui.adsabs.harvard.edu/abs/1977ApJ...217..693H} {217, 693}

\bibitem[\protect\citeauthoryear{{Herbig}, {Petrov}  \& {Duemmler}}{{Herbig}
  et~al.}{2003}]{Herbig03-FUOR}
{Herbig} G.~H.,  {Petrov} P.~P.,   {Duemmler} R.,  2003, \mn@doi [\apj]
  {10.1086/377194}, \href
  {https://ui.adsabs.harvard.edu/abs/2003ApJ...595..384H} {595, 384}

\bibitem[\protect\citeauthoryear{{Hirose}}{{Hirose}}{2015}]{Hirose15}
{Hirose} S.,  2015, \mn@doi [\mnras] {10.1093/mnras/stv203}, \href
  {https://ui.adsabs.harvard.edu/abs/2015MNRAS.448.3105H} {448, 3105}

\bibitem[\protect\citeauthoryear{{Lasota}}{{Lasota}}{2001}]{Lasota01-Review}
{Lasota} J.-P.,  2001, \mn@doi [\nar] {10.1016/S1387-6473(01)00112-9}, \href
  {https://ui.adsabs.harvard.edu/abs/2001NewAR..45..449L} {45, 449}

\bibitem[\protect\citeauthoryear{{Lodato} \& {Clarke}}{{Lodato} \&
  {Clarke}}{2004}]{LodatoClarke04}
{Lodato} G.,  {Clarke} C.~J.,  2004, \mn@doi [\mnras]
  {10.1111/j.1365-2966.2004.08112.x}, \href
  {http://adsabs.harvard.edu/abs/2004MNRAS.353..841L} {353, 841}

\bibitem[\protect\citeauthoryear{{Lodato}, {Nayakshin}, {King}  \&
  {Pringle}}{{Lodato} et~al.}{2009}]{LodatoEtal09}
{Lodato} G.,  {Nayakshin} S.,  {King} A.~R.,   {Pringle} J.~E.,  2009, \mn@doi
  [\mnras] {10.1111/j.1365-2966.2009.15179.x}, \href
  {http://adsabs.harvard.edu/abs/2009MNRAS.398.1392L} {398, 1392}

\bibitem[\protect\citeauthoryear{{Martin} \& {Lubow}}{{Martin} \&
  {Lubow}}{2014}]{Martin14-DZ-MRI}
{Martin} R.~G.,  {Lubow} S.~H.,  2014, \mn@doi [\mnras]
  {10.1093/mnras/stt1917}, \href
  {https://ui.adsabs.harvard.edu/abs/2014MNRAS.437..682M} {437, 682}

\bibitem[\protect\citeauthoryear{{Meyer} \& {Meyer-Hofmeister}}{{Meyer} \&
  {Meyer-Hofmeister}}{1984}]{Meyer84-CVs}
{Meyer} F.,  {Meyer-Hofmeister} E.,  1984, \aap, \href
  {https://ui.adsabs.harvard.edu/abs/1984A&A...132..143M} {132, 143}

\bibitem[\protect\citeauthoryear{{Nayakshin} \& {Elbakyan}}{{Nayakshin} \&
  {Elbakyan}}{2024}]{Nayakshin23-FUORi-2}
{Nayakshin} S.,  {Elbakyan} V.,  2024, \mn@doi [\mnras]
  {10.1093/mnras/stae049}, \href
  {https://ui.adsabs.harvard.edu/abs/2024MNRAS.528.2182N} {528, 2182}

\bibitem[\protect\citeauthoryear{{Nayakshin}, {Owen}  \&
  {Elbakyan}}{{Nayakshin} et~al.}{2023}]{Nayakshin-23-FUOR}
{Nayakshin} S.,  {Owen} J.~E.,   {Elbakyan} V.,  2023, \mn@doi [\mnras]
  {10.1093/mnras/stad1392}, \href
  {https://ui.adsabs.harvard.edu/abs/2023MNRAS.tmp.1409N} {}

\bibitem[\protect\citeauthoryear{{Nayakshin}, {S{\'a}enz de Miera},
  {K{\'o}sp{\'a}l}, {{\'C}alovi{\'c}}, {Eisl{\"o}ffel}  \& {Lin}}{{Nayakshin}
  et~al.}{2024}]{Nayakshin-24-classic-TI}
{Nayakshin} S.,  {S{\'a}enz de Miera} F.~C.,  {K{\'o}sp{\'a}l} {\'A}.,
  {{\'C}alovi{\'c}} A.,  {Eisl{\"o}ffel} J.,   {Lin} D. N.~C.,  2024, \mn@doi
  [\mnras] {10.1093/mnras/stae877}, \href
  {https://ui.adsabs.harvard.edu/abs/2024MNRAS.tmp..914N} {}

\bibitem[\protect\citeauthoryear{{Pavlyuchenkov}, {Akimkin}, {Topchieva}  \&
  {Vorobyov}}{{Pavlyuchenkov} et~al.}{2023}]{Pavlyuchenkov-23-TI}
{Pavlyuchenkov} Y.~N.,  {Akimkin} V.~V.,  {Topchieva} A.~P.,   {Vorobyov}
  E.~I.,  2023, \mn@doi [Astronomy Reports] {10.1134/S1063772923050086}, \href
  {https://ui.adsabs.harvard.edu/abs/2023ARep...67..470P} {67, 470}

\bibitem[\protect\citeauthoryear{{Powell}, {Irwin}, {Bouvier}  \&
  {Clarke}}{{Powell} et~al.}{2012}]{PowellEtal12}
{Powell} S.~L.,  {Irwin} M.,  {Bouvier} J.,   {Clarke} C.~J.,  2012, \mn@doi
  [\mnras] {10.1111/j.1365-2966.2012.21898.x}, \href
  {http://adsabs.harvard.edu/abs/2012MNRAS.426.3315P} {426, 3315}

\bibitem[\protect\citeauthoryear{{Scepi}, {Lesur}, {Dubus}  \& {Flock}}{{Scepi}
  et~al.}{2018}]{Scepi-18-TI-alpha}
{Scepi} N.,  {Lesur} G.,  {Dubus} G.,   {Flock} M.,  2018, \mn@doi [\aap]
  {10.1051/0004-6361/201731900}, \href
  {https://ui.adsabs.harvard.edu/abs/2018A&A...609A..77S} {609, A77}

\bibitem[\protect\citeauthoryear{{Scepi}, {Dubus}  \& {Lesur}}{{Scepi}
  et~al.}{2019}]{Scepi-19-DNe}
{Scepi} N.,  {Dubus} G.,   {Lesur} G.,  2019, \mn@doi [\aap]
  {10.1051/0004-6361/201834781}, \href
  {https://ui.adsabs.harvard.edu/abs/2019A&A...626A.116S} {626, A116}

\bibitem[\protect\citeauthoryear{{Shakura} \& {Sunyaev}}{{Shakura} \&
  {Sunyaev}}{1973}]{Shakura73}
{Shakura} N.~I.,  {Sunyaev} R.~A.,  1973, \aap, \href
  {http://cdsads.u-strasbg.fr/cgi-bin/nph-bib_query?bibcode=1973A%26A....24..337S&db_key=AST}
  {24, 337}

\bibitem[\protect\citeauthoryear{{Siwak} et~al.,}{{Siwak}
  et~al.}{2018}]{Siwak21-FUOri-QPOs}
{Siwak} M.,  et~al., 2018, \mn@doi [\aap] {10.1051/0004-6361/201833401}, \href
  {https://ui.adsabs.harvard.edu/abs/2018A&A...618A..79S} {618, A79}

\bibitem[\protect\citeauthoryear{{Smak}}{{Smak}}{1984}]{Smak84-dwarf-novae-review}
{Smak} J.,  1984, \actaa, \href
  {https://ui.adsabs.harvard.edu/abs/1984AcA....34..161S} {34, 161}

\bibitem[\protect\citeauthoryear{{Vorobyov} \& {Basu}}{{Vorobyov} \&
  {Basu}}{2006}]{VB06}
{Vorobyov} E.~I.,  {Basu} S.,  2006, \mn@doi [\apj] {10.1086/507320}, \href
  {http://adsabs.harvard.edu/abs/2006ApJ...650..956V} {650, 956}

\bibitem[\protect\citeauthoryear{{Vorobyov}, {Elbakyan}, {Liu}  \&
  {Takami}}{{Vorobyov} et~al.}{2021}]{Vorobyov-21-flyby-FUORs}
{Vorobyov} E.~I.,  {Elbakyan} V.~G.,  {Liu} H.~B.,   {Takami} M.,  2021,
  \mn@doi [\aap] {10.1051/0004-6361/202039391}, \href
  {https://ui.adsabs.harvard.edu/abs/2021A&A...647A..44V} {647, A44}

\bibitem[\protect\citeauthoryear{{Zhu}, {Espaillat}, {Hinkle}, {Hernandez},
  {Hartmann}  \& {Calvet}}{{Zhu} et~al.}{2009}]{Zhu09-FUOri-obs}
{Zhu} Z.,  {Espaillat} C.,  {Hinkle} K.,  {Hernandez} J.,  {Hartmann} L.,
  {Calvet} N.,  2009, \mn@doi [\apjl] {10.1088/0004-637X/694/1/L64}, \href
  {https://ui.adsabs.harvard.edu/abs/2009ApJ...694L..64Z} {694, L64}

\makeatother
\end{thebibliography}

\bsp	
\label{lastpage}
\end{document}